\documentclass[preprint,twocolum,amsmath,amssymb,aps,superscriptaddress]{revtex4-1}

\usepackage{graphicx}
\usepackage{dcolumn}
\usepackage{bm}
\usepackage{color}
\usepackage{multirow}
\usepackage{epstopdf}
\usepackage{ulem}
\usepackage[english]{babel}

\begin{document}

\preprint{APS/123-QED}

\title{{Optimizing the interfacial thermal conductance at gold-alkane junctions from `First Principles'}}

\author{Jingjie Zhang}
\email{jz9wp@virginia.edu}
\affiliation{Department of Electrical and Computer Engineering, University of Virginia, Charlottesville,VA-22904.}%

\author{Carlos A. Polanco}
\affiliation{Department of Electrical and Computer Engineering, University of Virginia, Charlottesville,VA-22904.}%

\author{Avik W. Ghosh}
\email{ag7rq@virginia.edu}
\affiliation{Department of Electrical and Computer Engineering, University of Virginia, Charlottesville,VA-22904.}%
\affiliation{Department of Physics, University of Virginia, Charlottesville,VA-22904.}

\date{\today}

\begin{abstract}
We theoretically explore the influence of end-group chemistry (bond stiffness and mass) on the interfacial thermal conductance at a gold-alkane interface. We accomplish this using the Non-Equilibrium Green's Function (NEGF) coupled with first principle parameters in Density Functional Theory (DFT) within the harmonic approximation. Our results indicate that the interfacial thermal conductance is not a monotonic function of either chemical parameters, but instead maximizes at an optimal set of mass and bonding strength. This maximum is a result of the interplay between the overlap in local density of states of the device and that in the contacts, as well as the phonon group velocity. We also demonstrate the intrinsic relationship between the Diffusive Mismatch Model (DMM) and the properties from NEGF, and provide an approach to get DMM from first principles NEGF. By comparing the NEGF based DMM conductance and range of conductance while altering the mass and bonding strength, we show that DMM provides an upper bound for elastic transport in this dimensionally mismatched system. We thus have a prescription to enhance the thermal conductance of systems at low temperatures or at low dimensions where inelastic scattering is considerably suppressed.
\end{abstract}

\pacs{Valid PACS appear here}

\maketitle

\section{Introduction}

Hybrid junctions between inorganic components and organic molecules have been of great interest due to their potential applications in electronic devices \cite{Nitzan2003,Joachim2000,Xu2003}, thermoelectrics \cite{Reddy2007,Baheti2008}, solar cells \cite{Tang2012,Ip2012} and optoelectronics \cite{Gather2011}. In recent years, the hybrid junctions, particularly the self-assembled monolayers (SAMs) have been intensively explored for thermal properties because of the discovery that the interfacial thermal conductance can be enhanced by more than a factor of four at room temperature simply by embedding the commonly assumed thermally insulating polymers (alkane chains) at the metal/dielectric interface \cite{OBrien2013a, Losego2012a}. This enhancement can even be tuned over one order of magnitude by carefully varying the interfacial properties (bond stiffness and mass) with end-group chemistry \cite{OBrien2013a}.

The observations of the influence of the interfacial properties on the interfacial thermal conductance diverge in the existing literature. One set of articles \cite{Shen2011, Duda2011, Giri2014} reports that the conductance increases monotonically with the bonding strength. This observation is commonly predicted by molecular dynamics simulations of interfaces between solids. A contrary viewpoint \cite{Zhang2011, Chen2015, Saltonstall2013, Hopkins2009a, Carlos2013} is arrived at by lattice dynamics and Green's Function on one dimensional harmonic chains, showing that a maximum conductance exists when the interface properties help match the impedance from two sides. This maximum conductance can be obtained with an optimal interfacial bonding (the harmonic mean of the bonding from two sides) and the favored connecting mass (the geometric mean of the masses from two sides) \cite{Klemens108} like scattering treatment would be consistent with this viewpoint, since the transmission depends quadratically on the mass deviation, producing a sweet spot along the transmission graph.

The discrepancy can arise from two sources. The first is the presence of additional inelastic channels created by phonon-phonon interactions across interfaces. Note that the theoretical  observations showing a monotonically increasing conductance with interfacial bonding strength involve molecular dynamic simulations, where anharmonicity plays a significant role \cite{Shen2011, Duda2011, Chalopin2012}. On the contrary, the maximum conductance arises in the absence of phonon-phonon interaction, relevant at low temperatures and low dimensions \cite{Zhang2011, Chen2015, Saltonstall2013, Hopkins2009a}. A separate origin for the discrepancy is a dimensionality mismatch across the interface. A monotonic increase is found in 3D-3D interfaces (e.g. thin films) where the system can be decoupled into separate 1D chains for each transverse momentum. In contrast, 3D-1D interfaces such as molecular wires on substrates end up coupling the 1D modes strongly. The interfacial bonding strength needed to achieve maximum conductance will be unrealistically large in the former, 3D-3D case. 

Recent research shows that anharmonicity has limited influence on the metal-SAM systems, which are dominated by elastic transport of phonons across the metal-alkane junctions \cite{Hu2010a, Taylor2015, Segal2002,Meier2014, Majumdar2015}. Phonon interference in the metal-SAM systems was  predicted both theoretically \cite{Segal2002, Hu2010a} and experimentally \cite{Meier2014}, demonstrating that phonons across the chains encounter only weak phonon-phonon interaction. S. Majumdar \cite{Majumdar2015} recently reported experimentally that the interfacial thermal conductance of this system decreases with the increased mismatch in frequency spectra between the two metal contacts, in fact behaving opposite to molecular dynamics simulations. This observation further validates elastic phonon transport along the chains. 

Motivated by the tuning capability of surface chemistry on the metal-alkane interfacial thermal conductance and the fact that phonon-phonon interaction in this system is trivial, we use ab-initio Non-Equilibrium Green's Function(NEGF)  method under harmonic approximation to explore the influence of the bond stiffness and end-group mass on the interfacial thermal conductance of the Au-alkane junctions. We show that the maximum conductance can be obtained at a set of optimal interfacial properties. This result clarifies that in a dimension mismatched heterojunction, even in the case that bulk contact serves as a reservoir of channels, matched impedance still improves the conductance. We generalize our results with a simple one degree of freedom 3D-1D model and show that the conductance can be tracked by the density of states(DOS), but is also influenced by the phonon group velocity.

The rest of the paper is organized as follows: First, we briefly discuss the structures of Au-alkane surfaces with both thiol and amine as end groups. Then we describe the way the bond stiffness is varied in this study. After that we introduce the first principle NEGF method with the "scooping" technique for the dimensionality transition followed by a short decription of NEGF based DMM method. Finally we discuss and analyze our first principle interfacial thermal conductance results and generalize them by using a simple 3D-1D model.

\section{Model and simulation method}
\subsection{Au/alkane surface structure}

\begin{figure}[ht]
	\centering
	\includegraphics[width=100mm]{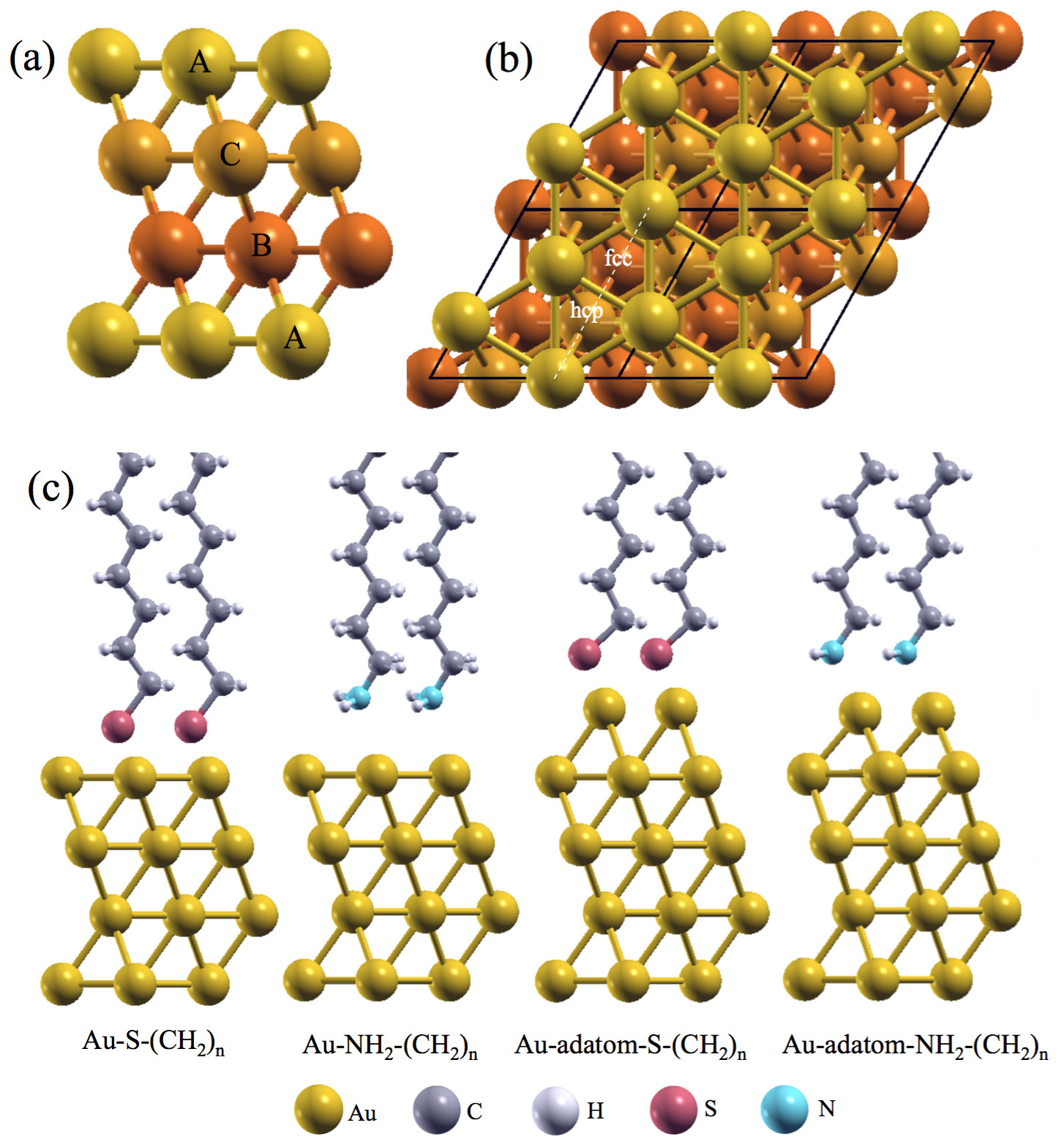}
	\caption{ (a) Stacking order of Au slab in (111) direction, the color difference represents the gold atoms in difference layers; (b) top view of the gold (111) surface, the black lines highlight the simulation supercell; (c) schematic representation of the systems being considered.}
	\label{Fig1structure}
\end{figure}

We model the Au-alkane surface using the $\sqrt{3}\times\sqrt{3}$ $R30^{o}$ \cite{Yourdshahyan2002, Vericat2010} unit cell. This structure corresponds to the highest density possible of alkane chains assembled on Au surface, which also leads to the highly ordered chains \cite{Vericat2010}. Fig.~\ref{Fig1structure} shows the top view of the absorption position of alkanes on Au surface and the stacked view of the junctions. We use the hcp position as the adsorption position in this simulation (Fig.~\ref{Fig1structure}(b)). The alkane chains before geometry optimization are set to tilt along the hcp-fcc dash lines in Fig.~\ref{Fig1structure}(b), with tilt angles of $10^o$, $20^o$ and $30^o$. This geometry is very close to the optimized ones  reported in the existing literature ($20^o \sim 30^o$ between the tilted alkanes and the normal to the gold surface). During the first principle simulations, the alkane length is set to be 6 C atoms in the chain with the end group as thiol or amine .We also considered the reconstruction of the gold surface\cite{Hayashi2001,Vargas2001,Rodriguez2003, Tachibana2002}. In this case, a gold adatom is initially centered on top of the hcp position with the alkane end-group connected to it. The whole system is then relaxed during the optimization process. Both the clean surface and the adatom surface are simulated as shown in Fig.~\ref{Fig1structure}(c).

The geometry optimization and the force constants are calculated using the first principles package Quantum Espresso \cite{Giannozzi2009a}. Ultrasoft pseudopotentials with Local Density Approximation exchange-correlation function are used in the simulation. The cutoff energy for wave functions is calibrated for convergence with a minimum of 35 Ry, while the cutoff energy for charge density is set to be 300 Ry. The threshold for the convergence of ground state energy is set to be $1.0\times10^{-10}$Ry. 

\begin{figure}[htb]
	\centering
	\includegraphics[width=100mm]{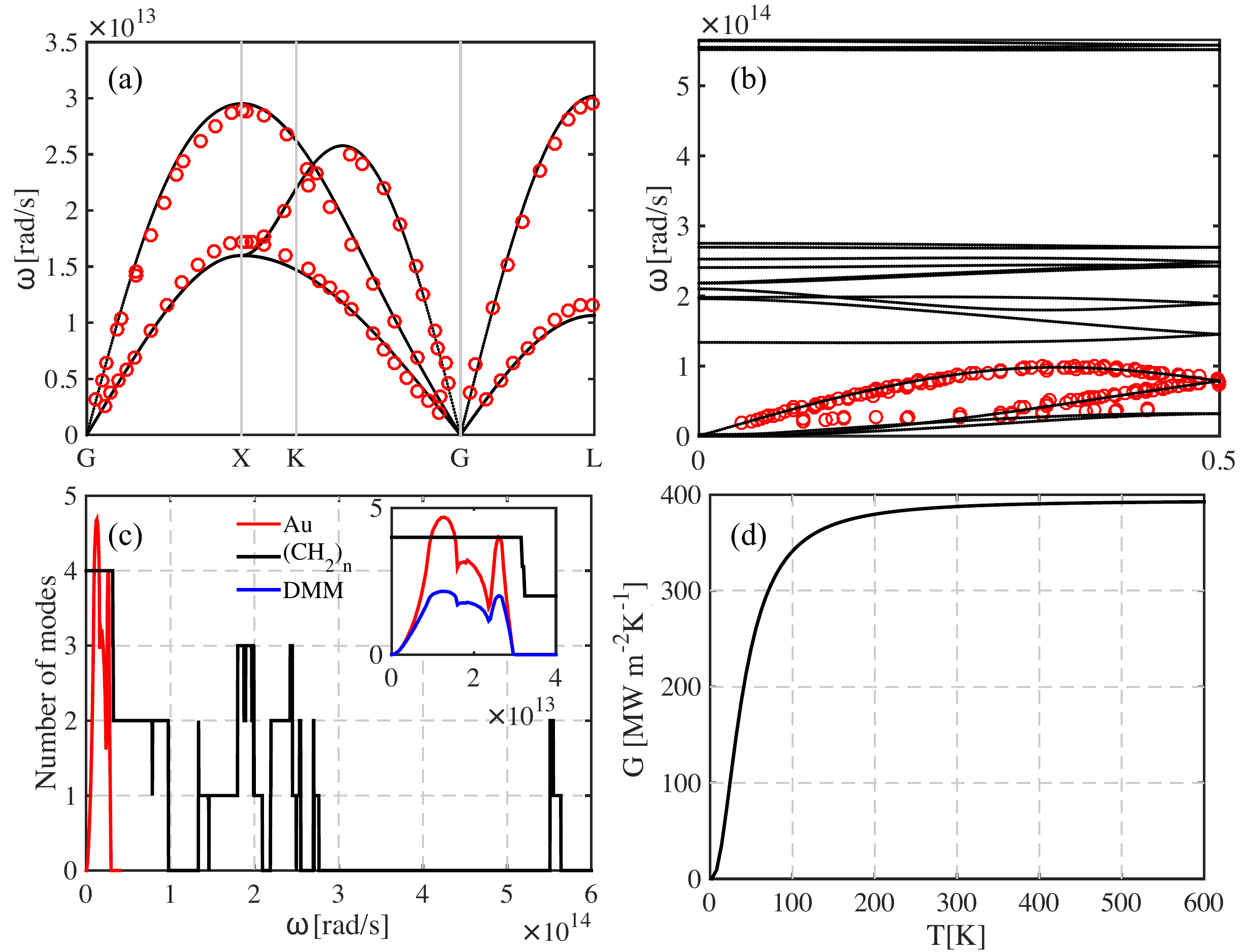}	 
	\caption{{\bf (a)} Dispersion of Au compared with experimental data; {\bf (b)} Dispersion of Polyethylene compared with experimental data; {\bf (c)} The number of modes of Au and polyethylene. The inset figure shows the MT of DMM taking the harmonic mean of modes from each side; {\bf (d)} The conductance of Au-SAMs junction with NEGF-based DMM.}
	\label{Fig2DMM}
\end{figure}

Before the geometry relaxation of the alkane chains assembled on the gold surface, we performed the structure optimization of bulk gold and infinite polyethylene. An  $8\times8\times8$ Monkhorst-Pack grid is used for k sampling in our bulk gold ground state simulation, with the methfessel-paxton smearing parameter set at 0.03Ry. A $4\times4\times4$ grid is used for q space sampling for gold force constants calculation using Density Functional Perturbation Theory (DFPT) \cite{Baroni2001}. In the polyethylene simulation, a $1\times1\times12$ grid is used in both the ground state calculation and DFPT in which the chain orientation is along the z direction. The distance between chains is set to be fairly large (10\AA) to avoid interactions between them. The optimized lattice constant for gold is 4.03 {\AA}  and the optimized length for one C$_2$H$_{4}$ unit is 2.5215 {\AA}. The dispersions using calculated force constants are shown in Fig.~\ref{Fig2DMM}(a) and Fig.~\ref{Fig2DMM}(b), along with the existing experimental data in the literature \cite{Lynn1973, DalCorso2013, Tomkinson2002a, Barrera2006}. 

The distance between the radicals and the gold surface before relaxation is set to be 2{\AA}. A supercell contains a $\sqrt{3}\times\sqrt{3}R30^{o}$ unit cell and four (111) Au layers is used as the model. 20{\AA} vacuum is set between the adsorbed alkane chains and the neighboring gold slab. A $3\times3\times1$ grid is used for both the structure optimization process and the force constant calculation process. The  threshold for force convergence is set to be $1.13\times10^{-3}$ Ry/{\AA}. 

After the calculation of the lattice dynamic matrix by DFPT, a Fourier transformation of the dynamic matrix is performed to obtain the real space force constant parameters. The q grid in DFPT calculations corresponds to the atomic interaction range. Up to 2nd neighboring atomic interactions are calculated for gold; up to 7th neighboring unit cell interactions are calculated for polyethylene. Although the $3\times3\times1$ grid provides the interaction between chains, in the following transport calculations, we ignore those interactions for simplicity.  
\subsection{NEGF and Landauer Formula}

\begin{figure}[htb]
	\centering
	\includegraphics[width=100mm]{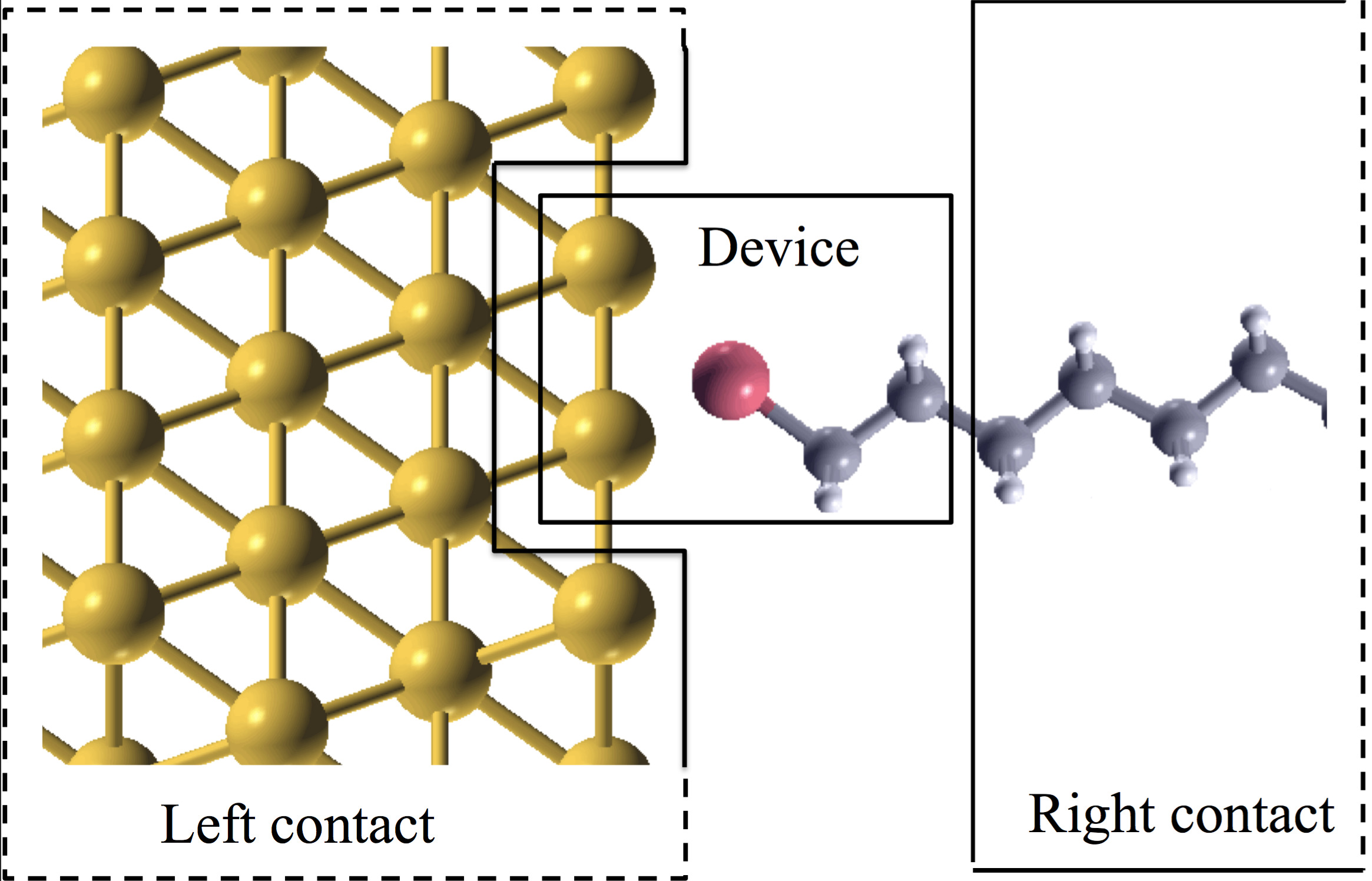}	
	\caption{ Schematic diagram of left, device and right regions of the alkane-gold junction in the NEGF simulation}
	\label{Fig3NEGF}
\end{figure}

In the Landauer description of phonon transport, the interfacial thermal conductance can be written as:
\begin{equation}
G=\int_0^{\infty}{\frac{\hbar\omega}{2\pi} M\bar{T} \frac{\partial N}{\partial T} dw} 
\label{Eq1Landauer}
\end{equation}

where $M$ is the number of modes and $\bar{T}$ is the mode-averaged transmission coefficient. $M\bar{T}$ can be calculated by the NEGF approach with the simulated system divided into left contact, right contact and a central device, as shown in Fig.~\ref{Fig3NEGF}: 
\begin{equation}
M\bar{T}=Trace[\Gamma_{l}G_d\Gamma_{r}G_d^{\dagger}]
\label{Eq2MT}
\end{equation} 

where $\Gamma_{l}$ and $\Gamma_{r}$ are the broadening matrices for the left and right contact, and $G_d$ is the retarded Green's Function for the device:
\begin{equation}
G_d=(M_dw^2-K_d-\Sigma_{l}-\Sigma_{r})^{-1}
\label{Eq3Gd}
\end{equation}

In the above equation, $M_d$ and $K_d$ are the mass matrix and the force constant matrix for the device respectively, while $\Sigma_{l}$ and $\Sigma_{r}$ are the self-energies for the contacts whose anti-Hermitian parts give us the broadening matrices $\Gamma_{l,r}$. 

In this dimension mismatched system, we invoke an additional step that we call "scooping" \cite{Zhang2007}. As shown in Fig.~\ref{Fig3NEGF},  the device contains 3 gold atoms around the hcp position (in the case of reconstructed gold surface with adatom, the scooped area contains 4 gold atoms with the adatom gold atom sitting on top of the triangle) in the gold surface layer, the end-group and the end C$_2$H$_4$ unit of the alkane.  The left contact is the semi-infinite bulk gold with the adsorption surface scooped out of it, while the right contact continues as the semi-infinite alkane chain.  Based on the optimized tilt angle, we rotate the force constants of alkanes to minimize the mismatch at the right contact.  The self-energy of the right contact is then evaluated as: 
\begin{equation}
\Sigma_{r}=K_{dr}g_rK_{rd}
\label{Eq4sigma}
\end{equation}

where $g_r$ is the surface green's function of the right contact. For the left contact with the scooped area, the self-energy can be calculated by:
\begin{equation}
\Sigma_{l}=M_{sp}w^2-K_{sp}-\widetilde{g_0}^{-1}
\label{Eq5sigma_r}
\end{equation}

with $M_{sp}$ being the mass matrix for the scooped area, and $K_{sp}$ the renormalized force constants following the acoustic sum rule(ASR) for the bare gold surface, with $\widetilde{g_0}$ the bare surface green's function of the scooped area. 
\subsection{DMM from ab-initio NEGF }

The Diffusive mismatch model (DMM) is commonly used to predict the thermal boundary conductance. In order to set a reference conductance for Au-SAM junctions, we calculate the DMM conductance from the first principle parameters. Instead of using the frequency dependent density of states and velocity from the full non-equilibrium lattice dynamics, we develope a quick estimate using DMM-NEGF, using variables calculated from a Green's function with {\it{ab-initio}} parameters. 

We start with the description of heat current across the interface in Landauer formula as in Eq.~\ref{Eq1Landauer}:
\begin{equation}
q^{1 \to 2}=\int_{0}^{\infty}(\frac{\hbar \omega}{2\pi}M_{1}\bar{T}^{1\to 2})\Delta Ndw
\label{Eq61to2}
\end{equation}

\begin{equation}
q^{2 \to 1}=\int_{0}^{\infty}(\frac{\hbar \omega}{2\pi}M_{2}\bar{T}^{2\to 1})\Delta Ndw
\label{Eq72to1}
\end{equation}

The DMM equations arise under the assumption of detailed balance\cite{Swartz1989, Duda2010}, the heat so that the current from material 1 to material 2 is equal to the current from material 2 to material 1:
\begin{equation}
q^{1 \to 2}=q^{2\to 1} \Rightarrow M_{1}\bar{T}^{1\to 2}=M_{2}\bar{T}^{2\to 1}
\label{Eq81to2equal2to1}
\end{equation}

Diffusive process of phonons across the interface assumes that the phonons randomize their phases at one-shot at the interface, consequently the transmission probability of phonons from material 1 to material 2 should equal to the reflection probability of phonons from material 2 to material 1:
\begin{equation}
\bar{T}^{1\to 2}=1-\bar{T}^{2\to 1}
\label{Eq9RequalT}
\end{equation}

Combining the Eq.~\ref{Eq81to2equal2to1} and Eq.~\ref{Eq9RequalT}, we get:
\begin{equation}
M\bar{T}_{DMM}=M_{1}\bar{T}^{1\to 2}=M_{2}\bar{T}^{2\to 1}=\frac{M_{1}M_{2}}{M_{1}+M_{2}}
\label{DMM_M}
\end{equation}

$M_1$ and $M_2$ are the number of modes of the bulk materials at each side of the interfaces. We can get  $M_1$ and $M_2$ similarly in Green's function by defining the contacts and junction as the same materials, using $M\bar{T}=Trace[\Gamma_{l}G\Gamma_{r}G^{\dagger}]$. In the bulk materials, for each mode, transmission probability is 1, so $M\bar{T}$ is just the number of modes. 

In a Green's function description, the number of modes is proportional to $AD(\omega)v(w)$ \textcolor{blue}, where A is the transverse area of the simulated junction unit cell, $D(w)$ and $v(w)$ are the frequency dependent density of states and velocity respectively.  If we write the parallel combination of modes in terms of density of states and velocity, we can recover the general expression often used to describe the DMM from lattice dynamics. 
For the Au-SAM system, we need to consider the modes of Au and modes of SAMs in the same area. The DMM conductance can be obtained from the combined mode density in parallel:
\begin{equation}
G_{DMM}=\int_0^{\infty}{\frac{\hbar\omega}{2\pi} \frac{(M_{1}/A_1)(M_{2}/A_2)}{(M_{1}/A_1)+(M_{2}/A_2)}}\frac{\partial N}{\partial T}dw
\label{DMM_G}
\end{equation}

In our simulations, we use a compact Au-SAM density with the surface geometry as $\sqrt{3}\times\sqrt{3} R30^{o}$. Each alkane chain connects with 3 Au atoms in an area of $2.16\times10^{-19}m^2$. The simulated number of modes for Au and alkane chain in that area are shown in Fig.~\ref{Fig2DMM}(c), with the inset graph showing the $M\bar{T}_{DMM}$. The number of modes for the Au with that specific density is 3 times as the number of modes of Au in primitive unit cell in (111) direction. The $M\bar{T}_{DMM}$ equals the number of modes of materials at two sides in parallel, so it should set the upper limit reference if the number of modes at two sides are vastly different. In the case where the number of modes at two sides are similar, DMM sets up the lower limit conductance reference only if the interface is a well matched interface without any roughness and disorder. 

\begin{figure}[htb]
	\centering
	\includegraphics[width=86mm]{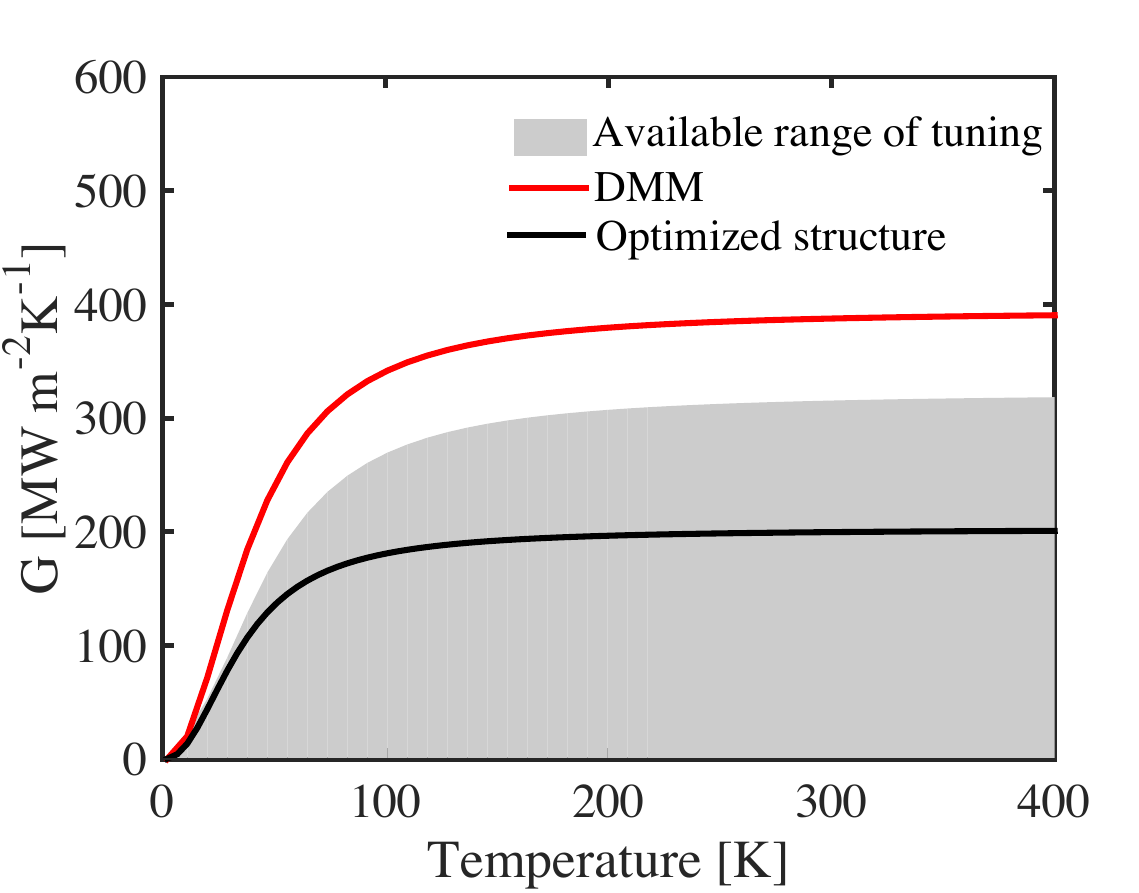}	
	\caption{ Tunable range of interfacial thermal conductance by varying the interface parameters compares to DMM conductance and the first principle optimized one. }
	\label{Fig4DMMrange}
\end{figure}

In the Au-SAM system, the dimensionality and property differ at the the interface when only gold atoms  are connected to the sulfur atom. Even though the number of modes of Au and alkane chain in the same area are similar to each other, the DMM conductance still sets up the upper bound to the interfacial conductance, as shown in Fig.~\ref{Fig4DMMrange}.

\section{ Influence of bonding and end group mass }\label{bonding mass}
\subsection{Comparison of different end groups}

Fig. 5 shows the force constants for different end groups obtained from first principles calculations, and the corresponding interfacial thermal conductance extracted from NEGF. The calculated trend of interfacial thermal conductances of different end-groups is consistent with the experimental data from Losego {\it{et al.}} \cite{Losego2012a}. The Au-thiol-alkane junctions have a larger interfacial bonding strength compared to Au-amine-alkane junctions, and a correspondingly larger interfacial thermal conductance. The interfacial thermal conductance follows the bonding strength in this case. However this is not conclusive because the bonding strengths are not large enough to attain maximum conductance.

The Au-thiol-alkane junction and Au-amine-alkane junction show different trends relating to the roughness of the gold surface. In the Au-amine-alkane junction, the reconstructed surface (adatom case) shows a larger conductance than that of the clean surface junction. In the Au-thiol-alkane junction,  the interfacial thermal conductance is larger for the clean gold surface. This discrepancy can be explained by analyzing the bond strength. The strength of all the bonds in all directions for Au-amine-alkane junction with adatom is larger than that of Au-amine-junction without adatom, hence the conductance is larger in the former case. In contrast, the reconstructed Au-alkanethiol junction has a bond only along the z direction is much larger than the clean surface Au case, which limits the coupling of transverse vibrational modes in gold to those within the alkane chain.

\begin{figure}[ht]
	\centering
	\includegraphics[width=100mm]{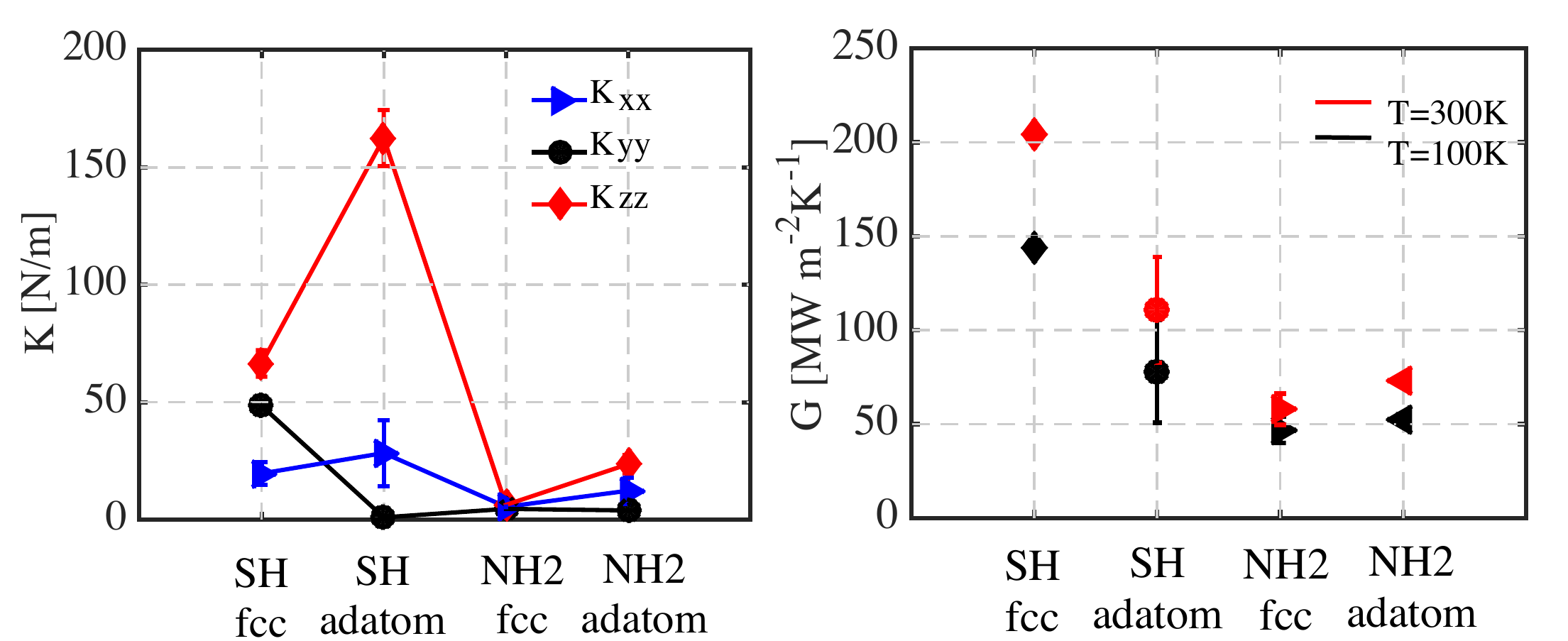}	
	\caption{The first principle NEGF result of the interfacial thermal conductance for the Au-thiol-alkane junctions and the Au-amine-alkane junctions. {\bf (a)} The bonding strength of different end groups on different adhesion sites. {\bf (b)} The corresponding interfacial thermal conductance.  The error bars show the range of conductance when the initial tilting angle is varied between $10^o$ and $30^o$. }
	\label{Fig5k}
\end{figure}

We note that the experimentally reported interfacial thermal conductance of a single Au-thiol-alkane junction is around 115 MWm$^{-2}$K$^{-1}$ \cite{Majumdar2015} and 220 MWm$^{-2}$K$^{-1}$ \cite{Wang2008}, while our results are around 110 MWm$^{-2}$K$^{-1}$ to 210 MWm$^{-2}$K$^{-1}$,  depending on details of the gold surface reconstruction. The calculated results are close to the previous experimental work, especially that of the clean gold surface compared to Wang et al. for the same system \cite{Wang2008}. 
\subsection{Role of interfacial parameters}

We use two different mechanisms to vary the interfacial bonding strength: varying the adsorption distance, or directly varying the bonding energy $\lambda$ relative to the Au-alkane thiol junction. In the case of varying adsorption distance, we assume that the alkane chains are normal to the gold surface, with the S atom centered on top the equilateral triangle formed by 3 gold atoms. In this step, the structure is not optimized during the ground state or during the lattice dynamic matrix calculations. The force constants are calculated by the small displacement method \cite{Kunc1982, Parlinski1997, Sluiter1999} by only moving the atoms around the interfaces. Alternately, we directly vary the interfacial bonding strength between the gold and the end-group relative to its calculated bond strength at an adsorption distance of 2{\AA} by a ratio of $\lambda$. For adjusting the end-group mass, we assume once again that the bond stiffness is fixed at its 2{\AA} value, and then change the corresponding mass matrix of the end-group. 

In order to focus on the influence of bonding strength and mass of the end-group attached to gold, we assume there is no impedance mismatch between the alkane end C$_2$H$_4$ unit and the thiol end-group, and set their bonding strength equal to  the one between two C$_2$H$_4$ units. This would be satisfied by enforcing the acoustic sum rule for the Sulfur atom. 

\begin{figure}[ht]
	\centering
	\includegraphics[width=86mm]{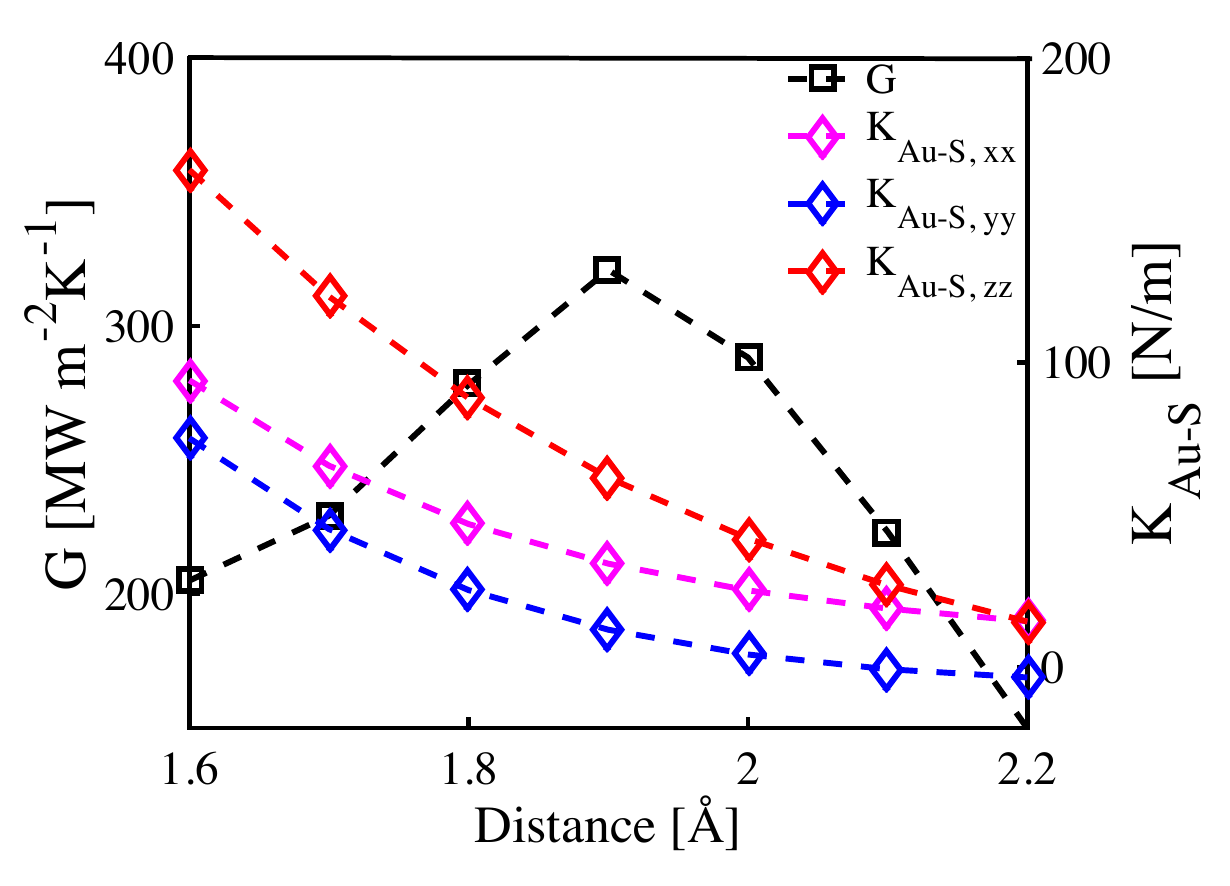}	
	\caption{The stiffness of the bonding varied with adhesion distance, and the corresponding thermal conductance of Au-SAMs junction.}
	\label{Fig6stress}
\end{figure}

\begin{figure}[ht]
	\centering
	\includegraphics[width=100mm]{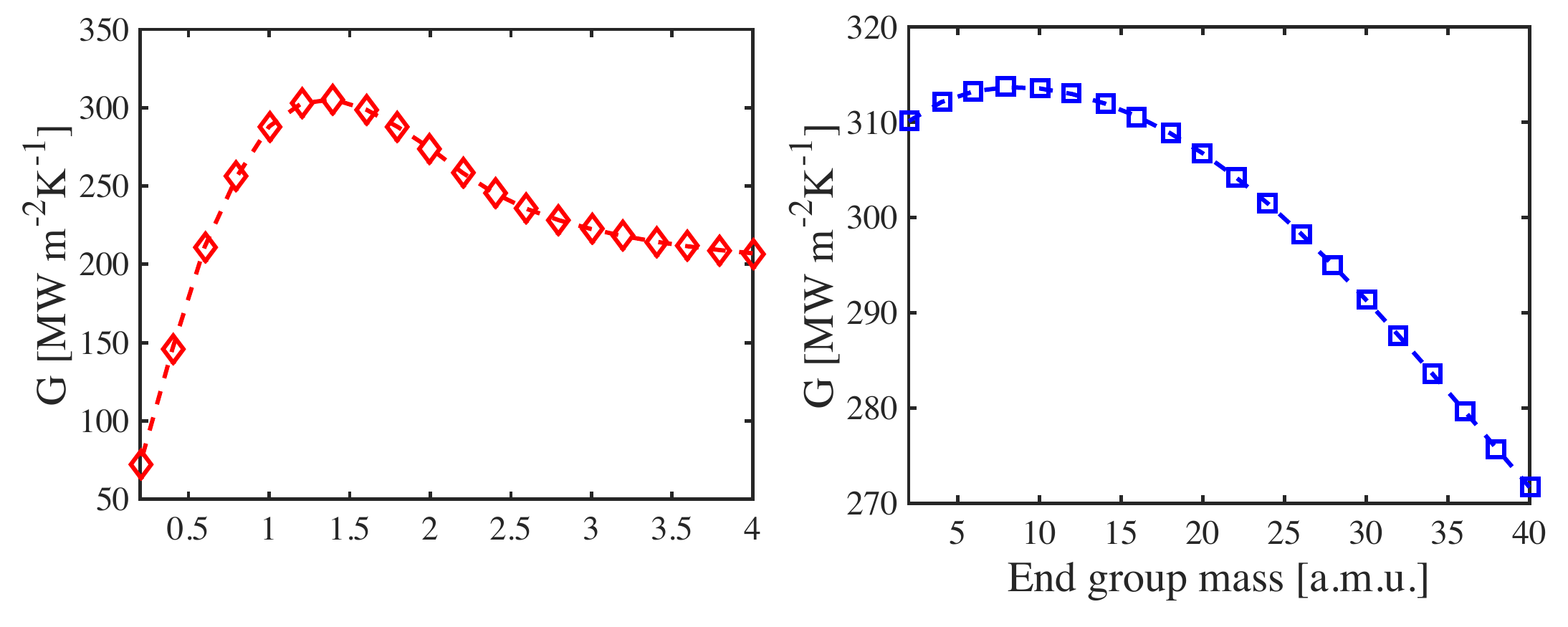}	
	\caption{Variation of the thermal conductance with directly changing {\bf (a)} the stiffness of the bonding and {\bf (b)} the mass of the adhesion species.  }
	\label{Fig7lambda}
\end{figure}

Fig.~\ref{Fig6stress} shows the variation of bond strength $K$ and thermal conductance $G$ with varying adsorption distance between 1.6$\AA$ and 2.2$\AA$. The interfacial bond strength shows an inverse proportionality to the adsorption distance. Curiously, the corresponding interfacial thermal conductance reaches a maximum at a distance 1.9{\AA}, which is close to the relaxed distance for alkanethiols self-assembled on gold. The existence of a maximum is independently validated when we directly vary the bond strength, as shown in Fig.~\ref{Fig7lambda}(a). A maximum conductance is seen to arise when the bond strength is around 1.3 times the initial 2$\AA$ distance value for thiol. When the bond strength continues to increase, the interfacial thermal conductance reduces and ultimately seems to saturate. 

A maximum is also observed around 10 a.m.u. when we plot the thermal conductance against the mass of the end group. This mass is lower than the mass of the C atom, indicating that the vibrational dynamics of the hydrogen atoms can not be ignored. For an impedance matched to maximize thermal transport across the interfaces, we would normally expect the preferred bonding mass to lie in between the masses of the two contacts. The maximizing mass also suggests that the prevailing Hautman-Klein  model \cite{Hautman1989} in molecular dynamics method used to simulate the alkane chains, where we neglect hydrogen atoms and add their masses instead to the carbon atoms, might not be suitable in the harmonic limit. 

We can use the NEGF extracted mode count in the DMM formula to directly estimate the thermal conductance (Fig.~\ref{Fig4DMMrange}), as discussed in the section III. The shaded area in this plot represents the tunable range of the interfacial thermal conductance reached by varying the interfacial strength and the end-group atomic mass. The range is predicted by the two approaches we described earlier in this section, which as we cautioned ignores the impedance between the alkane end C$_2$H$_4$ unit and the thiol anchoring group. Ignoring this contribution implies that the upper-limit of the tunable range could well be smaller than the shaded area in the harmonic assumption. The solid black line represents the contribution from a clean gold surface optimally bonded to an alkanethiol. Comparing with the tunable range, we can see that, in the harmonic limit the interfacial thermal conductance can at most be enhanced by less than $30\%$. This maximum is still less than the conductance that DMM predicts for this system. Since DMM arises in the limit of complete dephasing, it seems that such dephasing events must break restrictive symmetry selection rules that would otherwise limit the bandwidth of the transmitting phonon modes across the two dissimilar materials. In reality, we expect the answer to be slightly lower than DMM, because true dephasing events retain partial memory, and also tends to reduce the average transmission per mode.
\subsection{ 3D-1D generalized results and discussions }\label{simple 3D-1D}

\begin{figure}[ht]
	\centering
	\includegraphics[width=100mm]{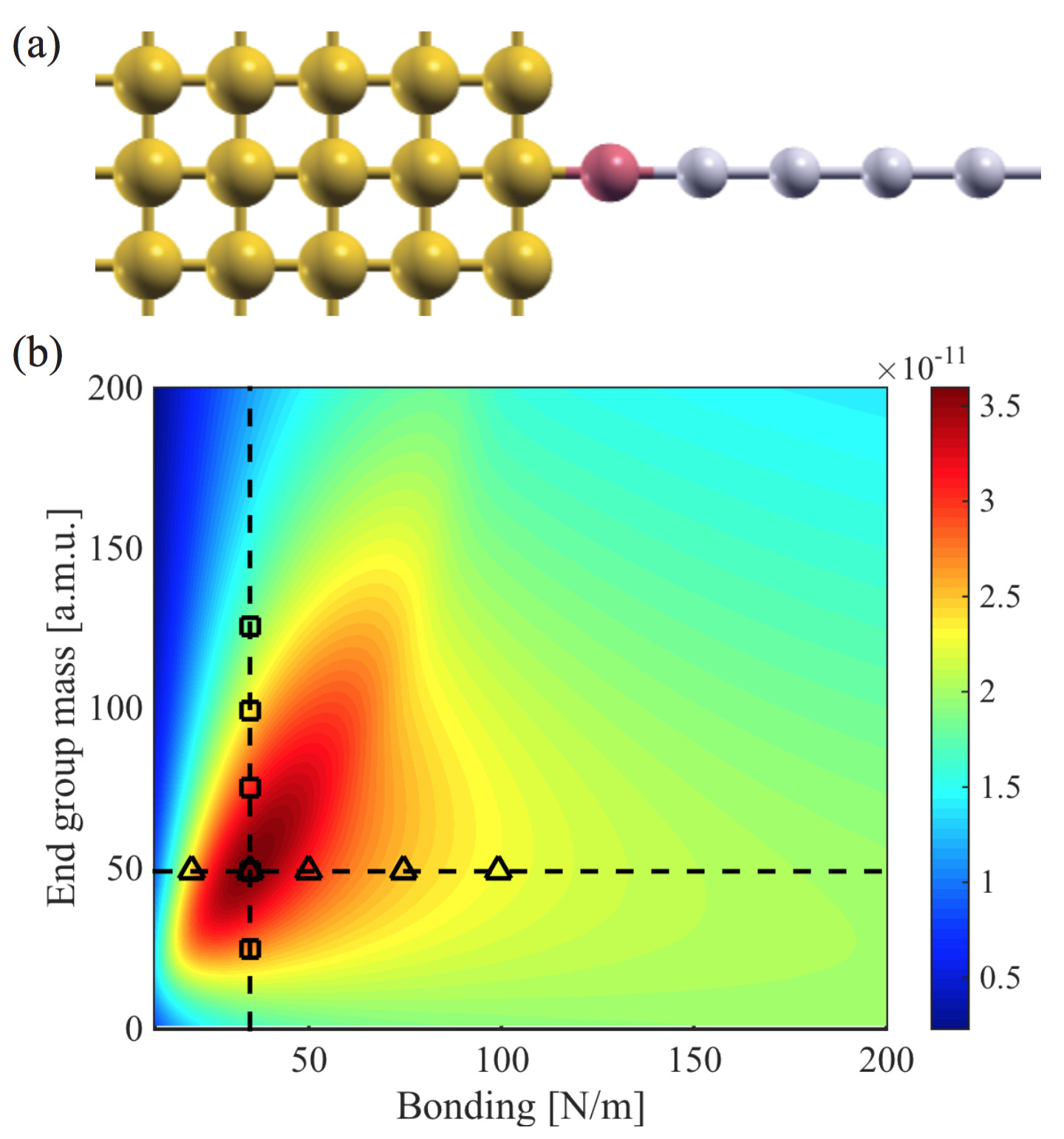}	
	\caption{ (a)A structural sketch of the simple 3D-1D model; (b)The thermal conductance (in unit [W/K]) map with respect to end group mass and stiffness of the interfacial bonding in the simple 3D-1D model.}
	\label{Fig8map}
\end{figure}

To explore whether our conclusions generalize to dimensional mismatch, we also studied a simplified one degree of freedom 3D-1D interface with only one degree of mechanical freedom (Fig.~\ref{Fig8map}(a)). In this model, the force constants exist in three dimensions to account for the transverse momenta in the 3D system, but the atoms are only allowed to move in the transport direction.  The mass of the atoms on the 3D side is set to that of gold, and the bond strength is set to match its cutoff frequency. Similarly, the atoms in the 1D chain have the mass of carbon while their bond strength matches the cutoff frequency of the alkane chain acoustic branches. The alkane channel is represented by a 1D channel between a 1D contact representing its extension on one side, and the 3D contact representing the Au substrate.

\begin{figure}[t]
	\centering
	\includegraphics[width=100mm]{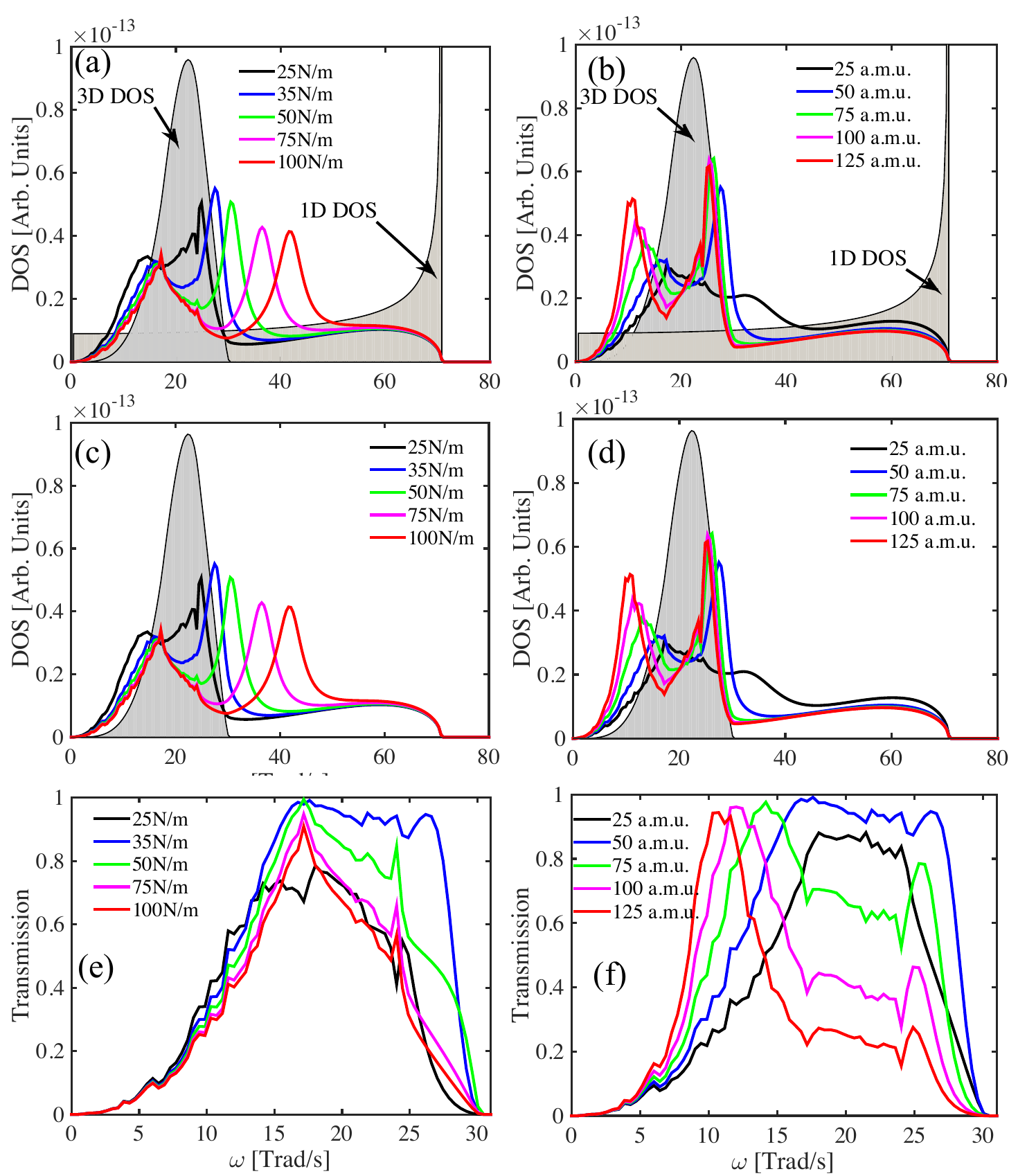}	
	\caption{ Density of states of the 3D contact, 1D contact and the device with different bonding strength {\bf (a)} and end group mass {\bf (b)} as shown in Fig.~\ref{Fig8map}(b) with square symbols and up triangle symbols respectively. The shadowed areas in {\bf(c)} and {\bf(d)} shows the averaged overlap density of states for the 3D and 1D contacts, dictating an density of states window for the DOS of the junctions. {\bf(e)} and {\bf(f)} shows the corresponding transmission for each set of interfacial parameters.}
	\label{Fig9simple3D1D}
\end{figure}

Fig.~\ref{Fig8map}(b) shows a map of the thermal conductance to visualize its dependence on interfacial parameters.The thermal conductance is scaled by the conductance in the classical limit. The figure confirms the existence of a maximum in the interfacial thermal conductance for a set of optimal interfacial parameters. We also see that the conductance is most sensitive to mass variations when the bond energy hits its maximizing value. In contrast, bond variations are critical when the endgroup masses are relatively high.

The existence of a maximum conductance with interfacial parameters can be explained by considering the dynamics of the density of states (DOS). Fig.~\ref{Fig9simple3D1D}(a), Fig.~\ref{Fig9simple3D1D}(c), Fig.~\ref{Fig9simple3D1D}(d) correspond to varying the bond strength with the end group mass set to 50 a.m.u. (up triangle symbols in Fig.~\ref{Fig8map}(b)), while Fig.~\ref{Fig9simple3D1D}(b), Fig.~\ref{Fig9simple3D1D}(d), and Fig.~\ref{Fig9simple3D1D}(e) correspond to varying the end group mass, with the bonding strength set to 35N/m (square symbols in Fig.~\ref{Fig8map}(b)). Let us first look at varying bonding strength. Fig.~\ref{Fig9simple3D1D}(a)  shows the DOS of the 3D and 1D contacts and the average local density of states (LDOS) of the device with various interfacial parameters. As the bonding strength between gold and end-group atom increases, the LDOS shifts towards the higher frequency states spanned by the 1D contact. This can be explained by the acoustic sum rule that once the bond strength increases, the on-site energy on the end-group atom will also increase, resulting in a up-shifted spectrum. These high frequency states cannot be used as transport states within a harmonic assumption, so that elastic transport cannot arise above the 3D cut-off frequency, in this case 30 Trad/s. 

The shaded area in Fig.~\ref{Fig9simple3D1D}(c) shows the overlap of DOS (ODOS) between the contacts :
\begin{equation}
ODOS(\omega) = \frac{D_{3D}(\omega)D_{1D}(\omega)}{\int_0^{\infty}D_{3D}(\omega)D_{1D}(\omega)d\omega}
\label{overlapDOS}
\end{equation}\\

where $D_{3D}(\omega)$ and $D_{1D}(\omega)$ are the DOS of 3D and 1D contact respectively. Within a harmonic assumption, the ODOS of contacts should serve as a weighted window for the channel states. As the carbon-carbon bonding strength increases, one of the LDOS peaks gets pushed out of the 3D band, resulting in an eventual decrease in transmission and its corresponding non-monotonicity with a preferred sweet spot. Comparing the DOS in Fig.~\ref{Fig9simple3D1D}(c) and transmission in Fig.~\ref{Fig9simple3D1D}(e), we see that the transmission approximately follows the LDOS of the devices within the weighted window. As described earlier, the difference between LDOS and transmission is the phonon group velocity. High frequency phonons have lower velocity, which explains the suppressed transmission of the LDOS peaks around the cutoff frequency.  Weaker bonding strength also create lower phonon velocity, so that the lowest transmission arises for the weakest bonding strengths.

The transmission dependence on end group mass has an origin similar to above, seen by comparing the transmission in Fig.~\ref{Fig9simple3D1D}(f) and the DOS in Fig.~\ref{Fig9simple3D1D}(d). Within the weighted window of the ODOS, the DOS for the lighter end group atom is much smaller, makings its transmission correspondingly weak. On the other hand, heavier end group mass reduces the phonon group velocity especially at higher frequency, reducing their corresponding transmissions as well. The end result is once again a maximum in the conductance when varying the mass of the end group atom.

\section{Conclusion}
In summary, using First principles phonon bands coupled with NEGF, we studied the influence of interfacial chemistry (bond strength and mass of the connector atoms) on the interfacial thermal conductance of gold-alkane junctions. Within a harmoonic approximation, we predict the existence of a maximum conductance that is mirrored by a simpler 3D-1D model. We attribute the maximum conductance to the interplay between local density of states and phonon group velocity in determining the phonon transmission. To estimate an upper limit to the conductance, we developed a DMM model based on the number of modes of the materials at each side of the interface extracted from the DFT-NEGF calculations.

\begin{acknowledgments}
J.Z., C.A.P. and A.W.G. are grateful for the support from NSF-CAREER (QMHP1028883) and from NSF-IDR (CBET 1134311). They also acknowledge the School of Engineering and Applied Sciences at University
of Virginia that covered the registration fees for a Quantum Espresso workshop. This work used the Extreme Science and Engineering Discovery Environment (XSEDE)(DMR130123) \cite{xsede}, which is supported by National Science Foundation grant number ACI-1053575.
\end{acknowledgments}



%

\end{document}